# Three-dimensional carbon nanotube networks from beta zeolite templates: Thermal stability and mechanical properties

Eliezer F. Oliveira[1,2,3*], Leonardo D. Machado[4], Ray H. Baughman[5], and Douglas S. Galvao[1,2*]

[1]Group of Organic Solids and New Materials (GSONM), Gleb Wataghin Institute of Physics, University of Campinas (UNICAMP), Campinas, SP, Brazil.

[2]Center for Computational Engineering & Sciences (CCES), University of Campinas (UNICAMP), Campinas, SP, Brazil.

[3]Department of Material Science and NanoEngineering, Rice University, Houston, Texas, 77005, United States.

[4]Department of Theoretical and Experimental Physics, Federal University of Rio Grande do Norte (UFRN), Natal, RN, Brazil.

[5]Alan G. MacDiarmid NanoTech Institute, The University of Texas at Dallas, Dallas, Texas, 75080-3021, United States.

*Corresponding authors: efoliver@ifi.unicamp.br, galvao@ifi.unicamp.br

**ABSTRACT**

We here investigated the thermal and mechanical behaviors of three-dimensional beta zeolite-templated carbon nanotube networks (BZCN). These networks are topologically generated by inserting carbon nanotubes (CNTs) into zeolite channels and connecting them using X-type junctions. We considered two cases, one with the tubes filling all zeolite channels (HD-BZCN) and the other with just partial filling (LD-BZCN). Fully atomistic reactive molecular dynamics (MD) simulations show that the networks exhibit high thermal stability (up to 1000 K). When compressed, the structures can withstand very large strains without fracturing (>50% for HD-BZCN and >70% for LD-BZCN). LD-BZCN can be stretched over 100% without fracturing.

# 1. INTRODUCTION

The search for new carbon-based nanostructures remains a very active research area [1-10]. These novel structures can have different dimensionalities (0D, 1D, 2D, and 3D) and exhibit a wide range of electrical, thermal, and mechanical properties [11, 12]. Of particular interest are 3D structures with well-ordered porous frameworks [13-15]. Previous studies predict that these carbon-based 3D networks display interesting electronic and mechanical properties, in addition to their large porosities and surface areas [13-16]. These properties could be exploited for a variety of applications in multiple technologies, including gas storage, catalysis, molecular sieving, and others [14, 15]. However, their synthesis has proved to be very challenging, especially for frameworks with covalently bonded building units [16].

One technique commonly employed to synthesize designed carbon-based structures is the use of sacrificial templates. Templating methods have been used since the 80s [17-19] to obtain carbon materials having different structural complexities [14]. Porous templates successfully used to obtain 3D carbon nanostructures include silica opals [20], metal-organic frameworks [21, 22], Ni foams [23], and mesoporous silica [24]. However, these 3D structures are in general disordered and/or with many structural defects [14-15]. In general, two major factors determine the quality of the resulting structures: (1) the method of synthesis and (2) the shape, length, and diameter of the template pores, which preclude or facilitate the diffusion of carbon atoms [14].

Recent studies have pointed out that it might be feasible to produce 3D carbon nanostructures with long-range order and few defects by using zeolites as sacrificial templates [13-15, 24, 25]. In general, zeolites are aluminosilicates containing molecular size pores and channels. These porous structures can enable the synthesis of 3D nanostructures within their interior [13, 14, 26]. In the most common synthesis methodology of zeolite-templated carbons (ZTC), carbon atoms are introduced into the zeolites via chemical vapor deposition (CVD) of carbon-containing precursors, and then the template is sacrificed/removed [14]. Currently, there are more than 200 types of zeolites [14, 27], although some are unsuitable for the production of covalently bonded 3D nanostructures, due to the absence of interconnected pores [14]. Also, it may be difficult to introduce precursors in zeolites having small pores, because of the difficulty of gas diffusion [15]. It has been suggested that only zeolites with pore sizes greater than 5 Å could be amenable for the experimental synthesis of ZTCs [15]. Consequently, for the synthesis of new 3D carbon-based nanostructures using zeolites, an important task is to properly screen zeolite databases to identify promising candidates.

Up to now, experimentally synthesized ZTCs consist of curved graphene nanosheets forming 3D structures [13, 14, 24, 28]. However, the synthesis of 3D carbon networks (CNs) of interconnected carbon nanotubes (CNTs) inside zeolite channels remains elusive [14] and only in the realm of theory [13, 14]. Such structures could be used for gas storage applications, as gas molecules strongly interact with curved surfaces [14-16, 29]. Theoretical studies have shown that CNT-based CNs present electronic and mechanical properties that are dominated by the network ordering [16]. Consequently, the properties of CNs could be tuned by engineering the network geometry [16]. In particular, charge transport calculations have shown that charges follow specific paths through the 3D structure, suggesting applications in nanoelectronics circuits [16].

The beta zeolite [27, 30], as an example shown in Figure 1, is an interesting candidate for CN synthesis, since this zeolite has equally spaced interconnected channels having the same diameter (~5.6 Å). This uniform channel size, which is not a general feature of zeolites, might allow the fabrication of a regular structure comprising fused nanotubes having the same diameter (see Figure 2).

We here investigate families of CNs that can be formed inside the channels of beta zeolite. Firstly, we carry out fully atomistic simulations to calculate the energy of different diameter nanotubes within the zeolite channels. Next, we create different 3D frameworks using selected CNTs. Finally, we examine the thermal stability and mechanical properties of the proposed structures.

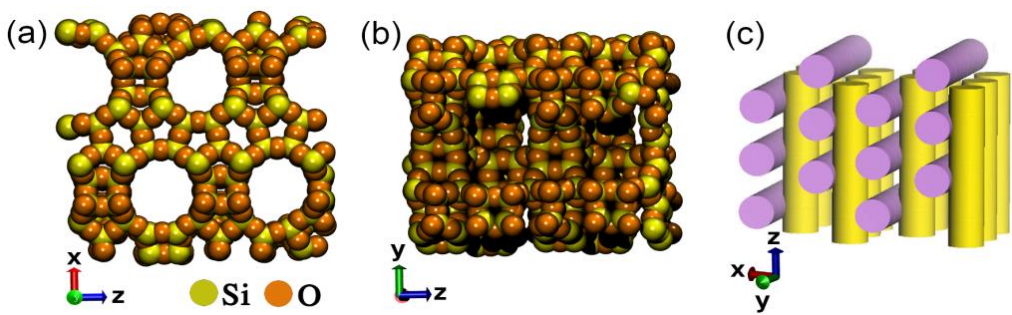

**Figure 1:** Structure of beta zeolite with BEA framework from (a) xz and (b) yz plane views. The channels along the y and z directions are orthogonal and partially cross each other. All channels are equivalent, with a diameter of 5.6 Å. (c) Representation of the channel disposition of beta

zeolite BEA. Cylinders in purple and yellow represents the channels along the y and z directions, respectively.

## 2. MATERIALS AND METHODS

For this study, a beta zeolite (BEA framework type) composed solely of silicon and oxygen atoms was used (see Figure 1). The BEA space group is P422, with unit cell parameters a=b=12.63 Å, c=26.18 Å, and α=β=ɣ=90° [30]. This zeolite has two orthogonal and equivalent channels (along the y and z directions, see Figure 1) that partially intersect. Figure 1(c) illustrates the relationship between the channels in BEA. In order to obtain single wall CNT candidates for forming the nets inside BEA, we determined which carbon nanotubes could fill the zeolite channels while providing a low energy structure. After selecting these CNT candidates, we then created energetically favorable junctions between the CNTs. The resulting building unit composed of interconnected nanotubes was then replicated to generate a periodic 3D structure. Finally, we removed the zeolite template so that the thermal stability and mechanical properties of the resulting CNs could be investigated. From here on we will use the acronym BZCN to refer to CNs templated from BEA.

To determine which CNT are most energetically suitable for filling the channels of BEA, we carried out Molecular Mechanics (MM) simulations using the well-known Universal Force Field (UFF) [31], as implemented in the Forcite software [32]. For these calculations, geometry optimization was performed with an energy convergence tolerance of 0.001 kcal/mol and a force convergence tolerance of 0.5 kcal/mol/Å. The size and shape of the unit cell was also optimized. This procedure was repeated for single wall nanotubes having different diameters. The used BEA unit cell contains 3840 atoms. The total number of atoms in the tested carbon nanotubes depended on their diameter, but the tube length was kept fixed (10 nm). We used a large periodic cell (12.3 nm) along the CNT axis direction, in order to prevent spurious inter-nanotube interactions.

After examining the CNT candidates, the next step was to generate junctions among the nanotubes. Two nanotubes were inserted into adjacent BEA channels, along the y and z directions, respectively (see Figure 1). The position of these nanotubes was fixed, and they were removed from the zeolite. In the region where the nanotubes were close to each other, atoms from both tubes were removed and placed into the space between them. Molecular dynamics (MD) simulations were then performed to create the junction, using the ReaxFF force field [33], as implemented in the computational package LAMMPS [34]. In these simulations, the temperature was controlled using a chain of three Nosé-Hoover thermostats, and a time step of 0.2 fs was used. During these simulations, the positions of the nanotubes was kept fixed and various cycles of minimization/heating/cooling (from 300 K to 1500 K) were performed to anneal defects in the junction region, until only energetically favorable pentagons, hexagons, and heptagons of carbon remained. It should be stressed that transverse channels in BEA do not intersect, and the nanotubes are connected by a "X" type junction, comprised of heptagons and pentagons [16], as displayed in Figure 2.

By inserting the covalently connected nanotubes back inside BEA, we determined if the created junction would fit inside the void that initially existed between CNTs. This process was used to create unit cells that could be replicated to generate a periodic 3D structure of arbitrary size.

After these steps, we carried out MD simulations to analyze thermal stability and mechanical properties of BZCN structures. After removing the zeolite, we minimized the energy of each BZCN using the conjugate gradient technique, and then equilibrated the system in an NPT ensemble with T = 300 K and P = 0 GPa and periodic boundary condition (PBC). Next, we applied either tensile or compressive uniaxial strain to deform the structure and evaluated its mechanical response to beyond the fracture limit. We used strain rates of $10^{-6}$ fs$^{-1}$ and $-10^{-6}$ fs$^{-1}$ for tension and compression, respectively. The deformation tests were performed at 300 K, and the pressure was set to 0 GPa along directions perpendicular to the external strain. The virial stress tensor component and the engineering strain at each compressive/tensile direction was used for building the stress-strain curve. The local stress distribution was evaluated through the von Mises stress analysis, using the second invariant of the deviatoric stress tensor [35]. This analysis allows us to determine where stress is spatially concentrated during compression/tensile tests, providing insights into the fracture dynamics. All deformation tests were performed using the ReaxFF force field [33], with a time step of 0.25 fs. The overall methodology applied here has been successfully used to study other carbon nanostructures [8, 9, 36].

## 3. RESULTS AND DISCUSSIONS

### 3.1. Obtaining BZCNs:

Initially, we inserted CNTs having different chiral indices (n,m) into BEA to determine the CNT with the lowest energy per carbon atom. The used unit cells contain seven channels, with a single nanotube embedded into the central channel (see Figure 2(a)). Figure 2(b) provides energy values *versus* nanotube diameter for (n, m) nanotubes. To obtain these energy values, first we determined the energy of the zeolite + nanotube structure, then we subtracted the energy of the isolated zeolite, and finally we divided the result by the number of carbon atoms. The results presented in Figure 2(b) show that (3,3), (4,2), (5,1), (6,0), (4,3), and (5,2) CNTs provide similar energy values. The minimum energy nanotube was the (6,0), which was used to design our carbon networks. Although we have not considered BZCN structures composed of nanotubes with mixed chiralities, it is important to remark that these structures are likely relevant, given that energies inside BEA are quite similar for a range of chiral indices.

BZCNs were created by adding multiple (6,0) carbon nanotubes to the BEA channels and then connecting them with "X" type junctions (Figure 2(c)) [37]. We considered two cases, with total and partial filling of BEA channels (hereafter name high density and low-density nets, respectively). These structures are shown in Figure 3. In Figures 3(a) and 3(b), we present unit cells that can be replicated to generate the BZCN. Hereafter, we use HD-BZCN/LD-BZCN to refer to high- and low-density frameworks, respectively. The unit cell of HD-BZCN has 321 carbon atoms (a=13.4 Å, b=13.4 Å, c=29.3 Å, and $\alpha=\beta=\gamma=90°$) and LD-BZCN has 894 carbon atoms (a=39.5 Å, b=39.5 Å, c=28.5 Å, and $\alpha=\beta=\gamma=90°$).

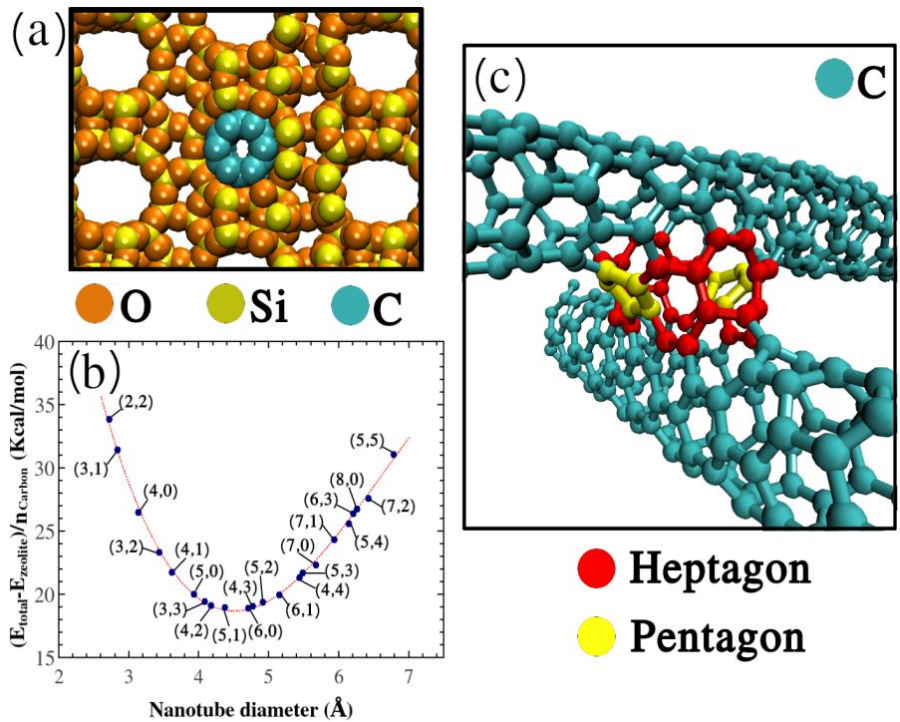

**Figure 2:** (a) Representative MD snapshot of a (6,0) carbon nanotube inside a BEA channel. (b) Energy per carbon atom for a CNT inserted into BEA *versus* nanotube diameter. $E_{total}$ is the total energy of a zeolite + nanotube system, $E_{zeolite}$ is the energy of an isolated zeolite, and $n_{Carbon}$ is the number of carbon atoms in a given nanotube. (c) An "X" type junction [37] between two (6,0) CNTs. Carbons in hexagons, heptagons and pentagons are indicated by green, red, and yellow spheres, respectively.

In order to verify whether both 3D networks remain stable inside BEA, we performed 30000 MD equilibration steps at 300K followed by 30000 steps at 600 K. After verifying that they remained stable, we removed the zeolite and minimized the energy of the isolated carbon networks, to confirm the stability of the obtained net structures. Both BZCN are less dense than other usual 3D carbon allotropes such as graphite (~2.20 g/cm3) and diamond (~3.50 g/cm3) [11]. The calculated density values for HD-BZCN and LD-BZCN are 1.28 g/cm3 and 0.44 g/cm3, respectively.

Figure 4 provides MD snapshots showing the porosity HD-BZCN and LD-BZCN. For HD-BZCN (Figure 4(a)), almost circular-like channels were observed along the xy plane, with a diameter of approximately 9.0 Å (cross sectional area of ~64.0 Å$_2$). For the xz plane, the channels are almost rectangular, with a cross sectional area of ~72.0 Å$_2$ (~8.0 Å x 9.0 Å). Finally, the channels for the yz plane are very narrow, with a diameter of ~4 Å (cross sectional area of ~12.5 Å$_2$).

The corresponding results for LD-BZCN are presented in Figure 4(b). The shapes of the channels are very similar for the xy and xz planes; these rectangular channels have a cross-sectional area of ~351.0 Å$_2$ (~9 Å X 39.0 Å). For the yz plane, there are differently sized channels, namely: (i) small square channels with a cross-sectional area of ~132.2 Å$_2$ (~11.6 Å x 11.4 Å); (ii) rectangular channels with a cross-sectional area of ~182 Å$_2$ (~11.5 Å x 15.8 Å); and (iii) large square channels with a cross-sectional area of 251.2 Å$_2$ (15.9 Å x 15.8 Å). Considering these results, these structures could potentially be used for gas storage, molecules and fast ion and molecular diffusion.

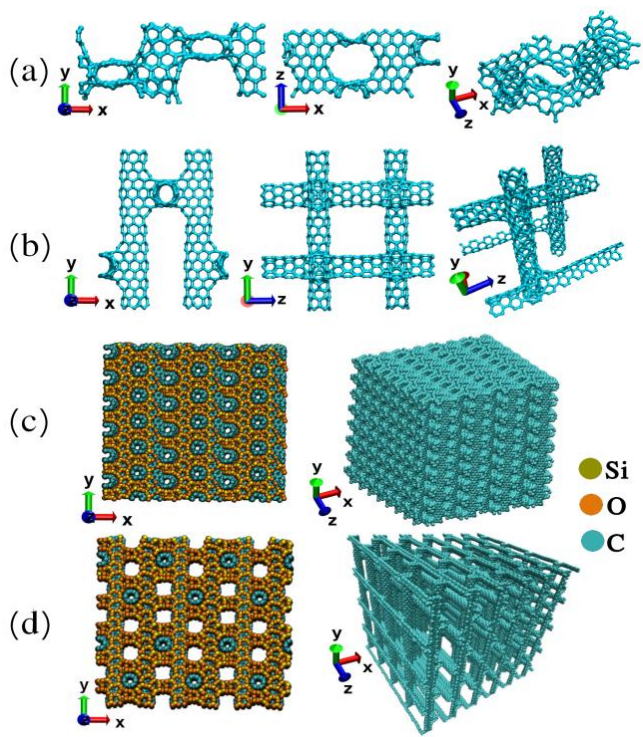

**Figure 3:** Different views of BZCN unit cells for (6,0) CNTs, for high (a) and low (b) density (a) unit cells. These cells are composed of 321 (a) and 894 (b) carbon atoms. The multiple snapshots correspond to the same structure viewed from different orientations. (c) and (d) display the corresponding BZCN with and without the zeolite template.

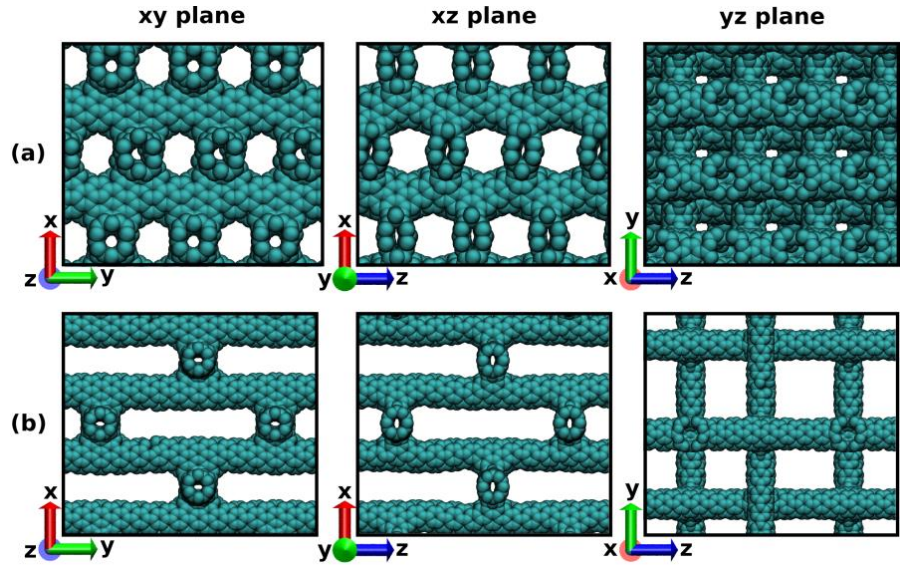

**Figure 4:** Different views of the channels in: (a) HD-BZCN and (b) LD-BZCN for (6,0) CNTs.

## 3.2. Mechanical properties of BZCNs:

We also investigated the mechanical behavior of BZCN under compressive and tensile strains. As the structures are porous and have large unit cells, it is important to determine whether the supercells used in the mechanical analyses are large enough to avoid spurious size-effects. For HD-BZCN, we tested structures with supercells of 1x2x2 (1284 atoms), 2x2x3 (3852 atoms), 2x3x3 (5778 atoms), 2x4x4 (10272 atoms), and 3x4x4 (15408 atoms). For LD-BZCN, we tested structures with supercells of 2x1x1 (1078 atoms), 2x2x1 (3576 atoms), 2x2x2 (7152 atoms), 3x2x2 (10728 atoms), and 3x3x2 (16092 atoms). For each structure, we first verified their stability at 300, 500, and 1000 K. All BZCN considered in these tests remained stable, even at high temperatures. Then, we applied to each structure a uniaxial tensile force along the y-direction (see Figure 3 for axis orientation). The results are presented in Figure 5. We obtained converged values from 10272 and 10728 atoms for HD-BZCN and LD-BZCN, respectively. Results are presented and discussed only for these structures.

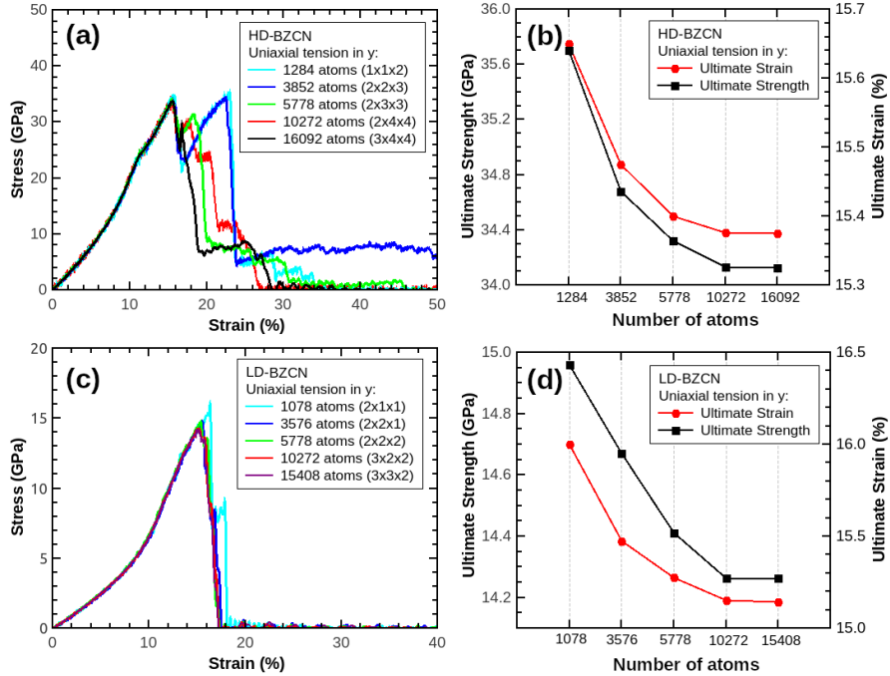

**Figure 5**: Mechanical properties for a tensile strain along the y-direction for HD-BZCN and LD-BZCN (6,0) CNT structures having different sizes. (a) Stress-strain curves and (b) the convergence of the ultimate strength and strain for HD-BZCN as a function of the number of atoms. Corresponding results are provided in (c) and (d) for LD-BZCN.

Figure 6 presents results for compressive and tensile strains (the entire deformation process can be seen in videos 1-8 in Supplementary Materials). Figures 6(a) and 6(b) provide results for HD-BZCN, and show its anisotropic behavior. During compression (Figure 6(a)) the stress remained almost constant (~10 GPa) from ~10% to ~40% strain. Interestingly, the stress level per atom remained well distributed throughout this strain range, as evidenced in the snapshots in Figures 7(a) and (b). This stress distribution can be attributed to the fact that deformation remained largely elastic for such large strain values. For larger strain values, stress rapidly increased (Figure 6(a)), and its distribution was no longer homogeneous, until fracture and amorphization occurred (Figures 7(a) and (b)).

Brittle behavior is observed in tensile strain simulations for HD-BZCN (Figure 6(b)). Also, the stress-strain curves are similar for the y and z directions, but rather different for the x direction.

This can be explained by considering how the CNTs are arranged along the three Cartesian directions. Figures 3 and 4 show that nanotubes are parallel to the y and z directions, whereas all nanotubes are perpendicular to the x direction. The von Mises stress distribution during uniaxial tensile deformation is presented in Figures 7(c) and (d). For all directions, the stress is spatially well distributed for low strain values (< 8%). For intermediate strain values (~13%), the stress is more equally dispersed along the x-direction, although with higher values at the junctions, as expected. For the y direction, stress accumulates in the nanotubes that are parallel to the tensile direction, and failure occurs when these CNTs break (Figure 7(d)). For the x-direction, failure occurs not at the junctions, but at the nanotubes that are perpendicular to the tensile stress direction (see Figure 7(c)). Results for the z direction are not displayed in this figure, because both the stress arrangement and the fracture process are quite similar in the y and z orientations.

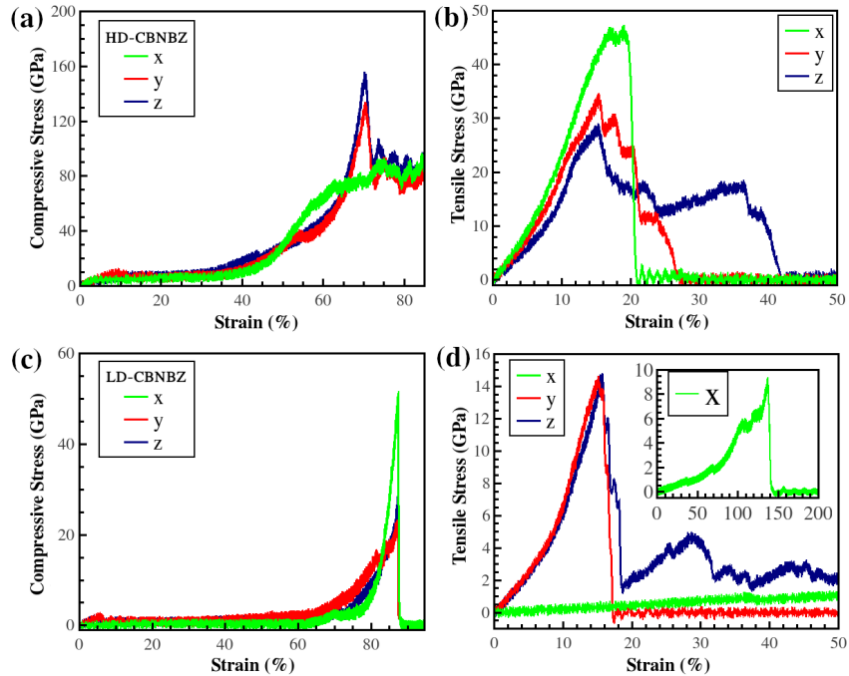

**Figure 6**: (a) Compression and (b) tensile stress-strain curves for HD-BZCN. Molecular dynamics simulations with the strain applied along the x, y, and z directions are indicated by using different color curves. (c) Compression and (d) tensile stress-strain curves for LD-BZCN. The inset in (d) shows the complete stress-strain curve for the x direction, where the tensile stress on the y-axis is in GPa and the percent strain is on the x-axis.

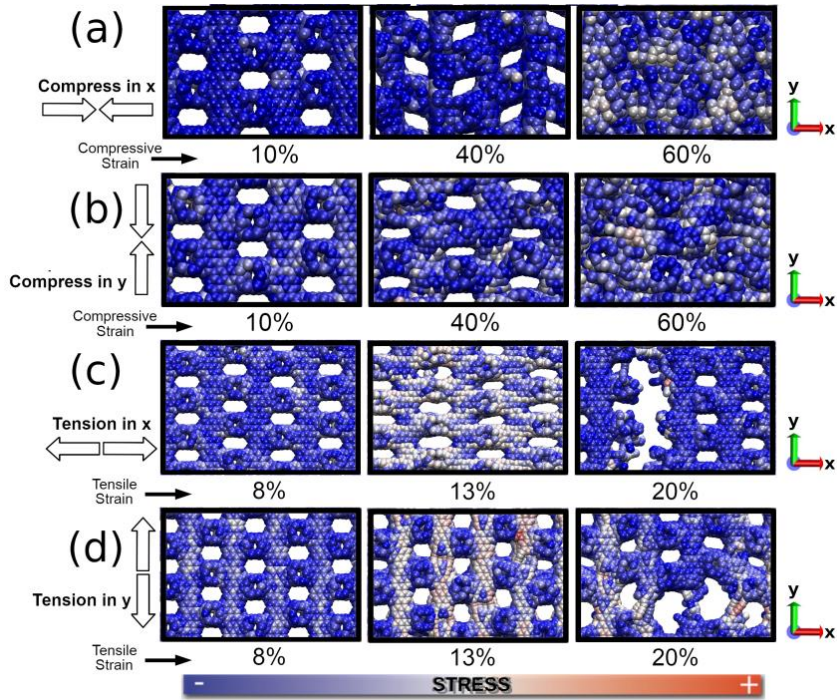

**Figure 7:** Representative MD snapshots for different compressive (a and b) and tensile (c and d) strains in the x and y directions of HD-BZCN. The color indicates the per atom von Mises stress value. As deformation along the y and z directions yield similar stress distributions, results for the z orientation were omitted.

Figures 6(c) and 6(d) provide corresponding results for LD-BZCN. The stress-strain curves for compression (Figure 6(c)) are similar in all directions for strains between ~5% and ~60%. For each direction, the stress to realize ~60% strain was low (no higher than ~3.0 GPa), and there was a relatively uniform distribution of local stress values (see Figures 8(a) and (b)). Amorphization only began when strain reached ~75% (see Figures 8(a) and (b)). It should be stressed that although large deformations without structural failures are commonly observed in non-crystalline foam-like materials, this behavior is rare for defectless single crystalline structures [38], which makes LD-BZCN a good candidate for applications that require large deformations without fracture.

The tensile stress-strain curves for LD-BZCN show that it is a brittle (except for stretch in the x-direction) and highly anisotropic (Figure 6(d)). The tensile stress-strain curve is quite similar when the strain is applied along the y or z direction, but quite different when it is applied along the x-direction. This behavior was also observed for HD-BZCN and can be explained by the same structural features: LD-BZCN has some nanotubes that are parallel to the y and z directions, while all CNTs are perpendicular to the x direction (Figure 4). For the y and z directions, the estimated values of the ultimate tensile strength (14.3 and 14.4 GPa, respectively) and strain (~15%) are similar. The results for the x-direction are more interesting. The stiffness is low, but very large strains can be applied. The strain value at the threshold of structural failure was 134%, and stress was ~9 GPa (see the inset of Figure 6(d)).

In Figures 8(c) and (d) we present representative MD snapshots detailing the von Mises stress values during the tensile process of LD-BZCN. For low strains, the structural stress is relatively well distributed for every direction (x,y, and z). As strain increases for the y and z directions, the stress is accumulated in the nanotubes that are parallel to the tensile direction, eventually leading to the structural failure of these CNTs (Figure 8(d)). For the x-direction, the deformation processes are more complex. As the strain increases, the nanotubes that were initially perpendicular to the tensile direction start to buckle, becoming increasingly aligned with the x-direction, as can be seen in Figure 8(c). For this orientation, the stress is initially concentrated in the junctions. Then, as the realignment process of nanotubes continues, the stress is concentrated in the CNTs, which eventually fail.

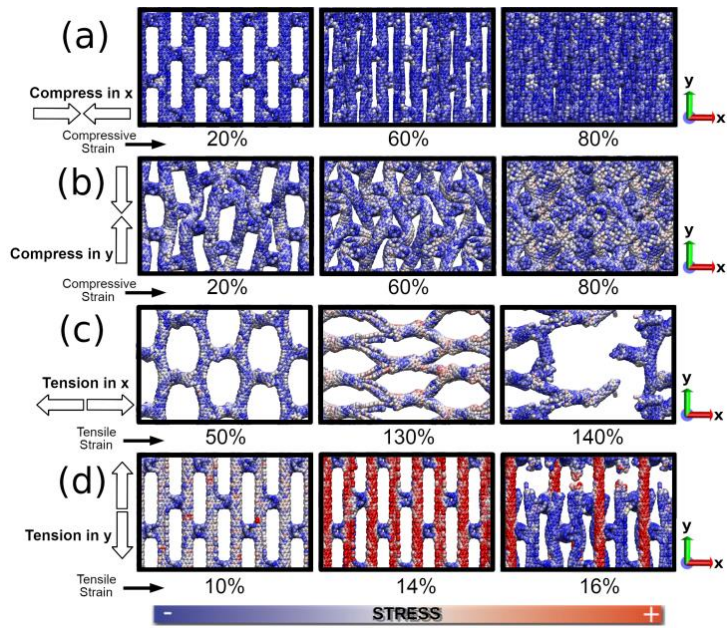

**Figure 8:** Representative MD snapshots for different compressive (a and b) and tensile (c and d) strains in the x and y directions of LD-BZCN. The color indicates the per atom von Mises stress

value. As deformation along the y and z directions yield similar stress distributions, results for the z orientation were omitted.

It is informative to compare HD-BZCN and LD-BZCN with other carbon allotropes. In Table 1 we present the ultimate strength, ultimate strain, and Young's modulus of HD-BZCN, LD-BZCN, diamond, graphene, and carbon nanotube. All results presented in this table were obtained using ReaxFF [36], following the same methodology. When compared to other carbon allotropes, HD-BZCN and LD-BZCN exhibit lower values of ultimate strength and Young's modulus. These results are expected, considering that these BZCN phases have porous structures. The calculated ultimate strengths for these BZCN phases are higher than for many other materials, such as silicon, steel, and titanium alloys [39]. The Young's modulus of HD-BZCN is in the range found for some metals, while LD-BZCN is much less stiff. Finally, the calculated ultimate strains are similar to the other carbon phases in Table 1, with the exception of the LD-BZCN x-direction, where we have the unusual result of low value for Young's modulus, but a very high value for the ultimate strain. As discussed above, this results from the CN topology.

**Table 1:** Predicted values for the ultimate strength, strain, and Young's modulus at room temperature for high density and low density BZCN. For comparison, the corresponding values for graphene, diamond, and nanotubes, which were obtained by using the same methodology [36], are also presented.

| Structure | Ultimate Strength (GPa) | Ultimate Strain (%) | Young's Modulus (GPa) |
|---|---|---|---|
| **HD-BZCN (x-direction)** | 46.2 | 16.8 | 157.0 |
| **HD-BZCN (y-direction)** | 34.1 | 15.3 | 143.6 |
| **HD-BZCN (z-direction)** | 28.3 | 15.4 | 109.8 |
| **LD-BZCN (x-direction)** | 9.1 | 134.1 | 1.5 |
| **LD-BZCN (y-direction)** | 14.3 | 15.2 | 44.9 |
| **LD-BZCN (z-direction)** | 14.4 | 15.8 | 41.5 |
| **Diamond** | 148.7 | 18.9 | 1300.0 |
| **Graphene (armchair)** | 149.6 | 21.5 | 1266.0 |
| **Graphene (zigzag)** | 115.4 | 26.1 | 1296.0 |
| **Nanotube (12,0)** | 111.3 | 21.6 | 1214.0 |

## 4. SUMMARY AND CONCLUSIONS

In this work, we investigated beta zeolite-templated nanotube-based 3D carbon networks (BZCNs). These networks are topologically generated by inserting carbon nanotubes (CNTs) into the zeolite channels and connecting them using X-type junctions, which are composed of only pentagons and heptagons. Our molecular mechanics calculations show there are many candidate tubes to generate the networks, with similar energy values. The results presented here are for the case of the (6,0) nanotube. We considered two cases, one with the tubes filling all zeolite channels (HD-BZCN) and one with partial filling (LD-BZCN).

We carried out fully atomistic reactive molecular dynamics (MD) simulations to investigate thermal stability and mechanical behavior under compressive and tensile loadings. Our results show that the networks exhibit high thermal stability (up to 1000 K). For compressive loadings, the structures were able to withstand very large strains (>50% for HD-BZCN and >70% for LD-BZCN) without fracture. With respect to tensile loadings, both structures exhibited brittle and anisotropic behaviors. The anisotropy is more significant for the lower density structure, with Young's modulus varying by a factor of ~30. For the low stiffness direction, the ultimate tensile strain value was very high, 134.1%.

Considering that there have been important advances in the zeolite-template synthesis [14,15], the production of large-size BZCN structures could be a reality in the coming years. Recently 3D printed versions of porous carbon-based models were reported [40] and surprisingly some mechanical behaviors proved to be scale independent. This could be also the case for BZCN. Works along these lines are in progress.

**ACKNOWLEDGMENTS**

We would like to thank the Brazilian agencies CNPq and FAPESP (Grants 2013/08293-7, 2016/18499-0, and 2019/07157-9) for financial support. Computational and financial support from the Center for Computational Engineering and Sciences at Unicamp through the FAPESP/CEPID Grant No. 2013/08293-7 is also acknowledged. This work was financed in part by the Coordenação de Aperfeiçoamento de Pessoal de Nível Superior - Brasil (CAPES) - Finance Code 001. LDM acknowledges the support of the High-Performance Computing Center at UFRN (NPAD/UFRN).

**REFERENCES**

[1] M. Terrones, A. Jorio, M. Endo, A. M. Rao, Y. A. Kim, T. Hayashi, H. Terrones, J. C. Charlier, G. Dresselhaus, and M. S. Dresselhaus, New direction in nanotube science, Mater. Today 7 (2004) 30-45.

[2] C. He, L. Sun, C. Zhang, J. Zhong, Two viable three-dimensional carbon semiconductors with an entirely $sp_2$ configuration, Phys. Chem. Chem. Phys. 15 (2013) 680-684.

[3] X.-L. Sheng, Q.-B. Yan, F. Ye, Q.-R. Zheng, G. Su, T-Carbon: A novel carbon allotrope, Phys. Rev. Lett. 106 (2011) 155703.

[4] L.A. Burchfield, M.A. Fahim, R.S. Wittman, F. Delodovici, N. Manini, Novamene: A new class of carbon allotropes, Heliyon 2 (2017) 3, e00242.

[5] N.V.R. Nulakani, V. Subramanian, Cp-Graphyne: A low-energy graphyne polymorph with double distorted Dirac points, ACS Omega 2 (2017) 6822-6830.

[6] F. Delodovici, N. Manini, R.S. Wittman, D.S. Choi, M.A. Fahim, L.A. Burchfield, Protomene: A new carbon allotrope, Carbon 126 (2018) 547-579.

[7] X. Wang, J. Rong, Y. Song, X. Yu, Z. Zhan, J. Deng, QPHT-graphene: A new two-dimensional metallic carbon allotrope, Phys. Lett. A 381 (2017) 2845-2849.

[8] E. F. Oliveira, P. A. S. Autreto, C. F. Woellner, D. S. Galvao, On the mechanical properties of novamene: A fully atomistic molecular dynamics and DFT investigation, Carbon 139 (2018) 782-788.

[9] E. F. Oliveira, P. A. S. Autreto, C. F. Woellner, D. S. Galvao, On the mechanical properties of protomene: A theoretical investigation, Comput. Mat. Sci. 161 (2019) 190-198.

[10] Q. Wei, Q. Zhang, H. Yan, M. Zhang, B. Wei, A new tetragonal superhard metallic carbon allotrope, J. Alloy Compd. 769 (2018) 347-352.

[11] T.D. Burchell, Carbon Materials for Advanced Technologies, first ed., Elsevier Science, Oxford, 1999.

[12] G. Messina, S. Santangelo, Carbon: the Future Material for Advanced Technology Applications, first ed., Springer, New York, 2006.

[13] R.H.Baughman, D.S.Galvao, Tubulanes: carbon phases based on cross-linked fullerene tubules, Chem. Phys. Lett. 221 (1993) 110.

[14] H. Nishihara, T. Kyotani, Zeolite-templated carbons – three-dimensional microporous graphene frameworks, Chem. Commun. 54 (2018) 5648.

[15] E. Braun, Y. Lee, S. M. Moosavi, S. Barthel, R. Mercado, I. A. Baburin, D. M. Proserpio, B. Smit, Generating carbon schwarzites via zeolite-templating, PNAS 35 (2018) E8116-E8124.

[16] J. M. Romo-Herrera, M. Terrones, H. Terrones, S. Dag, V. Meunier, Covalent 2D and 3D networks from 1D nanostructures: Designing new materials, Nano Lett. 7 (2007) 570-576.

[17] M. T. Gilbert, J. H. Knox and B. Kaur, Porous glassy carbon, a new columns packing material for gas chromatography and high-performance liquid chromatography, Chromatographia,16 (1982) 138–148.

[18] R. W. Pekala and R. W. Hopper, Low-density microcellular carbon foams, J. Mater. Sci. 22 (1987) 1840–1844.

[19] T. Kyotani, N. Sonobe and A. Tomita, Formation of highly orientated graphite from polyacrylonitrile by using a two-dimensional space between montmorillonite lamellae, Nature 331 (1988) 331–333.

[20] A. A. Zakhidov, R. H. Baughman, Z. Iqbal, C. X. Cui, I. Khayrullin, S. O. Dantas, J. Marti, V. G. Ralchenko, Carbon structures with three-dimensional periodicity at optical wavelengths, Science 282 (1998) 897–901.

[21] B. Liu, H. Shioyama, T. Akita and Q. Xu, Metal-organic framework as a template for porous carbon synthesis, J. Am. Chem. Soc. 130 (2008 )5390–5391.

[22] L. Radhakrishnan, J. Reboul, S. Furukawa, P. Srinivasu, S. Kitagawa, Y. Yamauchi, Preparation of microporous carbon fibers through carbonization of Al-based porous coordination polymer (Al-PCP) with furfuryl alcohol, Chem. Mater. 23 (2011) 1225–1231.

[23] Z. P. Chen, W. C. Ren, L. B. Gao, B. L. Liu, S. F. Pei and H. M. Cheng, Three-dimensional flexible and conductive interconnected graphene networks grown by chemical vapour deposition, Nat. Mater. 10 (2011) 424–428.

[24] R. Ryoo, S. H. Joo and S. Jun, Synthesis of highly ordered carbon molecular sieves via template-mediated structural transformation, J. Phys. Chem. B 103 (1999) 7743–7746.

[25] K. Kim, T. Lee, Y. Kwon, Y. Seo, J. Song, J. K. Park, H. Lee, J. Y. Park, H. Ihee, S. J. Cho, R. Ryoo, Lanthanum-catalysed synthesis of microporous 3D graphene-like carbons in a zeolite template, Nature, 535 (2016) 131-135.

[26] G-H Moon, A. Bahr, H. Tuysuz, Structural engineering of 3D carbon materials from transition metal ion-exchanged Y zeolite templates, Chem. Mater. 30 (2018) 3779-3788.

[27] http://www.iza-online.org/

[28] H. Nishihara, Q-H. Yang, P-X. Hou, M. Unno, S. Yamauchi, R. Saito, Carbon 47 (2009) 1220.

[29] G. E. Froudakis, Mater. Today 14 (2011) 324.

[30] J.B.Higgins, R.B. LaPierrea, J.L. Schlenker, A.C. Rohrmana, J.D. Wooda, G.T. Kerr, and W.J. Rohrbaugha, Zeolites 8 (1988) 446-452.

[31] A.K. Rappe, C.J. Casewit, K.S. Colwell, W.A. Goddard III, W.M. Skiff, J.Am. Chem. Soc. 114 (1992) 10024–10035.

[32] http://lms.chem.tamu.edu/cerius2.html

[33] A.C.T. van Duin, S. Dasgupta, F. Lorant, W.A. Goddard, ReaxFF: A reactive force field for hydrocarbons, J. Phys. Chem. A 105 (41) (2001) 9396-9409.

[34] S.J. Plimpton, Fast parallel algorithms for short-range molecular dynamics, Comput. Phys. 117 (1995) 1-19.

[35] A. Zang, O. Stephansson, Stress Field of the Earth's Crust, first ed., Springer, Houten, 2009.

[36] B.D. Jensen, K.E. Wise, G.M. Odegard, Simulation of the elastic and ultimate tensile properties of diamond, graphene, carbon nanotubes, and amorphous carbon using a revised ReaxFF parametrization, J. Phys. Chem. A 119 (2015) 9710-9721.

[37] I. Zsoldos, G. Kakuk, T. Reti, A. Szasz, Geometric construction of carbon nanotube junctions, Modelling Simul. Mater. Sci. Eng. 12 (2004) 1251-1266.

[38] L. C. Felix, C. F. Woellner, D. S. Galvao, Mechanical and energy-absorption properties of schwarzites, arXiv:1909.03098v1 (2019).

[39] R.E. Smallman, A.H.W. Ngan, Physical Metallurgy and Advanced Materials Engineering, seventh ed., Elsevier, Butterworth-Heinemann, 2007.

[40] S. M. Sajadi, P. S. Owuor, S. Schara, C. F. Woellner, V. Rodrigues, R. Vajtai, J. Lou, D. S. Galvao, C. S. Tiwary, P. M. Ajayan, Multiscale Geometric Design Principles Applied to 3D Printed Schwarzites, Adv. Mater. 30 (2018) 1704820.